\documentclass[preprint,floatfix] {revtex4} 
\newcommand{\rvec}{\mathrm {\mathbf {r}}} 
\newcommand{\rpvec}{\mathrm{\mathbf{r'}}}
\usepackage{graphicx}
\usepackage{subfigure}
\usepackage{longtable}
\begin{document}

\title{Ground and excited states of Li$^-$, Be$^-$ through a density-based approach}
\author{Amlan K. Roy}
\altaffiliation{Corresponding author. Email: akroy@chem.ucla.edu. Present address: Department of 
Chemistry, University of Kansas, Lawrence, KS, 66045, USA.}
\affiliation{Department of Chemistry and Biochemistry, University of California, 
Los Angeles, CA, 90095-1569, USA}

\author{Abraham F. Jalbout}
\affiliation{Institute of Chemistry, National Autonomous University of Mexico,
Mexico City, Mexico}

\begin{abstract}
Density functional calculations are performed for ground [He]2s$^2$ $^1$S$^e$, and three metastable 
bound excited states, 1s2s2p$^2$ $^5$P$^e$, 1s2p$^3$ $^5$S$^o$, 1s2s2p3p $^5$P$^e$ of Li$^-$ 
and [He]2s2p$^2$ $^4$P$^e$, [He]2p$^3$ $^4$S$^o$, 1s2s2p$^3$ $^6$S$^o$ of Be$^-$ each. The 
work-function-based exchange potential is used, while the correlation effects are included by 
employing the Lee-Yang-Parr potential. The relevant nonrelativistic KS equation is solved 
by means of a generalized pseudospectral discretization scheme offering nonuniform and optimal 
spatial grid. Computed total energies, radial densities, selected density moments, as well as two 
transition wavelengths (1s2s2p$^2$ $^5$P$^e \!\! \rightarrow$1s2p$^3$ $^5$S$^o$ of Li$^-$, 
[He]2s2p$^2$ $^4$P$^e\!\! \rightarrow$ [He]2p$^3$ $^4$S$^o$ of Be$^-$) show reasonably good 
agreement with the available theoretical and experimental data. The term energies show an absolute 
deviation of 0.007--0.171\% with the largest deviation being observed for the even-parity $^5$P 
state of Li$^-$. The transition wavelengths of Li$^-$, Be$^-$ are calculated within 0.891 and 
0.438\% of the experimental values. This offers a simple practical route towards accurate reliable 
calculation of excited states of anions within density functional theory.
\end{abstract}
\maketitle

\section{Introduction}
Atomic negative ions are representative of weakly-bound (typically having binding energies one 
order of magnitude smaller than the neutral atoms or positive ions) fragile quantum systems. Due
to the weak central field experienced by its outer valence electron, correlation is dominant and
can no longer be treated as a small perturbation to the independent particle picture. Thus 
faithful estimation of binding energies, fine structure or excitation energies of negative ions
pose considerable challenges to theoreticians and experimentalists. For example, \emph{very good} 
precision on total energy is necessary to obtain \emph{modest} values of the electron affinity. An 
impressive amount of theoretical as well as experimental works have been reported in the literature
(see for example, \cite{buckman94,andersen97} for review) during the past three decades and interest
in this area continues to grow further. 
 
Our focus is on the ground and excited states of negative ions of two most extensively studied 
systems (after possibly H$^-$ and He$^-$) from alkali and alkaline-earth groups, \emph{viz.,} Li, Be. 
Ever since the three long-lived 
bound states of Li$^-$ were reported through configuration-interaction (CI) calculations 
\cite{bunge80,bunge80a}, significant attention has been paid on their characterization and 
understanding. These three are respectively: the ground state [He]2s$^2$ $^1$S$^e$ and two 
core-excited high-spin metastable bound states, 1s2s2p$^2$ $^5$P$^e$, 1s2p$^3$ $^5$S$^o$, each lying 
below the corresponding 1s2s2p $^4$P and 1s2p$^2$ $^4$P parent states of Li. Thereafter a host of 
calculations including multiconfiguration Hartree-Fock (MCHF) \cite{fischer90}, multiconfiguration 
interaction \cite{yang95} as well as the recent saddle-point \cite{wang06}, variational Monte Carlo 
(VMC) \cite{galvez06}, and other methods have been employed to determine the upper bound, 
relativistic and nonrelativistic energy, transition wavelength, oscillator strength, lifetime, 
fine structure, hyperfine parameter, etc., of these states. In general, these findings show good 
agreement with the 
experimental observations \cite{berry70,mannervik80}. In contrast to Li$^-$ ground state, addition 
of a 2p electron to the ground state of Be 1s$^2$2s$^2$ yields an unbound Be$^-$ [He]2s$^2$2p 
$^2$P$^o$ ground state; however, three metastable states [He]2s2p$^2$ $^4$P$^e$, [He]2p$^3$ 
$^4$S$^o$ and 1s2s2p$^3$ $^6$S$^o$ were predicted, of which the first two have been observed 
experimentally \cite{kvale85,gaardsted89,kristensen95}. Theoretical calculations include a variety 
of methods, e.g., superposition of configurations \cite{weiss68}, CI \cite{beck81,bunge86}, 
state-specific theory \cite{aspromallis85}, MCHF \cite{fischer90,brage91,olsen94}, full core plus 
correlation \cite{hsu95}, Rayleigh-Ritz variation method \cite{hsu95a}, VMC \cite{galvez06}, etc.,
among others.

In the last four decades, density functional theory (DFT) \cite{parr89} has emerged as one of the 
most powerful and promising tools for electronic structure and dynamical studies of many-electron 
atoms, molecules, solids. While it enjoyed remarkable success for the ground states, same for 
excited states came much later. In the last decade, a work-function-based DFT prescription has 
been shown to be fairly successful for \emph{general} atomic excited states (see for 
example, \cite{roy97,roy98} and the references therein). This proposed the use of a nonvariational 
local work-function-based exchange potential \cite{harbola89}, which is computationally advantageous
compared to the nonlocal HF potential. Usually, the gradient and Laplacian-included correlation 
functional of Lee-Yang-Parr (LYP) \cite{lee88} was employed in conjunction. Recently this was extended 
for higher 
excitations through a generalized pseudospectral (GPS) implementation which allowed to solve the 
resultant radial KS equation in a non-uniform, optimal spatial grid accurately efficiently 
\cite{roy02}-\cite{roy05}. This enabled us to study, for the first time, within the framework of
DFT, a broad range of important physical processes in atomic excited states such as multiple 
excitations, 
valence as well as core excitations, satellite states, high-lying Rydberg states etc. However with 
the exception of 8 triply excited states of He$^-$ \cite{roy97}, all of these works dealt 
with either neutral atom and cations. The current work aims at investigating the spectra of 
two most important and widely studied case of negative ions through the above DFT method, with an 
objective to assess and judge the performance of this simple single-determinantal approach in the
context of excited states of negative ions. In other words can a density-based approach as 
ours, produce satisfactory and physically meaningful results for these intricate excited 
states; and if so how good do they perform in comparison to the traditional wave function-based
methodologies, which are prevalent for their studies? To answer these questions, we investigate the 
term energies, radial densities, transition wavelengths as well as expectation values of some 
position operators for 4 states of Li$^-$, \emph{viz.,} [He]2s$^2$ $^1$S$^e$, 1s2s2p$^2$ $^5$P$^e$, 
1s2p$^3$ $^5$S$^o$, 1s2s2p3p $^5$P$^e$, and 3 states of Be$^-$, \emph{viz.,} [He]2s2p$^2$ $^4$P$^e$, 
[He]2p$^3$ $^4$S$^o$ and 1s2s2p$^3$ $^6$S$^o$. Comparison with available 
theoretical and experimental results are made, wherever possible. The article is organized as 
follows: Section II gives basic elements of the formalism as well as its numerical implementation, 
Section III makes a discussion on the results while we end with a few concluding remarks in 
Section IV. 

\section{Method of calculation}
\label{sec:method}
The methodology as well as the numerical solution of the desired radial Kohn-Sham (KS) equation 
through GPS discretization scheme has been presented before [23,24,27-31]; hence not repeated. 
Only essentials details are given. Unless otherwise mentioned, atomic units employed 
throughout the article. 

Our interest is in the following single-particle nonrelativistic KS equation, 
\begin{equation}
\left[-\frac{1}{2} \nabla^2 + v_{es}(\rvec)
+v_{xc}(\rvec) \right] \psi_i (\rvec) = \varepsilon_i \psi_i (\rvec)
\end{equation}
where three terms in the left-hand side signify respectively the kinetic, electrostatic and
exchange-correlation (XC) energy contributions. $v_{es}(\rvec)$ contains the nuclear-attraction 
and classical internuclear Coulomb repulsion as,
\begin{equation}
\mathit{v}_{es}(\rvec)=-\frac{Z}{r} + \int \frac{\rho(\rpvec)}{|\rvec-\rpvec|} 
\mathrm{d}\rpvec
\end{equation}
where Z corresponds to the nuclear charge. The local nonvariational exchange potential 
relates to the work required to move an electron against the electric 
field $\boldmath{\cal{E}}_x(\rvec)$ arising out of its own Fermi-hole charge distribution,
$\rho_x(\rvec,\rpvec)$, and given by the line integral \cite{harbola89},
\begin{equation}
\mathit{v}_x (\rvec) = - \int_{\infty}^{r} 
\mbox{\boldmath $\cal{E}$}_x (\rvec) \cdot \mathrm{d} \mathbf{l}.
\end{equation}
where
\begin{equation} 
\mbox{\boldmath $\cal{E}$}_x(\rvec) = \int \frac 
{\rho_x (\rvec,\rpvec)(\rvec -\rpvec)} {|\rvec-\rpvec|^3} \ \ \mathrm{d}\rvec. 
\end{equation}
For well-defined potentials, work done must be path-independent (irrotational), which is 
rigorously satisfied for spherically symmetric systems. This potential can be 
calculated accurately as the Fermi hole is known exactly in terms of the single-particle orbitals.
Working within the central-field approximation, $\psi_i(\rvec)= R_{nl}(r)\ Y_{lm}(\Omega)$, 
employing a suitable correlation functional (here we use LYP potential, \cite{lee88}), solution of
the KS equation produces a self-consistent set of orbitals, which gives the electron density 
as, \[\rho(\rvec)=\sum_{\mathit{i}} |\psi_{\mathit{i}}(\rvec)|^2. \]

The radial KS equation is solved accurately and efficiently by means of the GPS method, which has
shown remarkable success for structure and dynamics of Coulombic singular systems such as atoms,
molecules as well as other difficult and stronger singularities such as the spiked harmonic 
oscillators, Hulthen and Yukawa potentials etc. \cite{roy02}-\cite{roy05}. In most of these 
cases, this has either outperformed the best results published so far for those systems or offered 
results of comparable accuracy to the best calculations. This is achieved through the following
steps: (i) approximate a function by an Nth order polynomial (we use Legendre) $f_N(x)$ in such a 
way that the approximation is \emph{exact} at collocation points $x_j$, i.e., $f_N(x_j)\! 
=\!f(x_j)$ (ii) a transformation $r \! = \! r(x)$ employed to map the semi-infinite domain $r \! 
\in \! [0,\infty]$ onto the finite domain $x \! \in \! [-1,1]$ (iii) introducing an algebraic 
nonlinear mapping $r \!= \! r(x) \! = \! L (1+x)/(1-x+\alpha)$, with $\alpha \! = \! 2L/r_{max}$, 
followed by a (iv) symmetrization procedure. Finally this leads to a \emph{symmetric} eigenvalue 
problem which can be easily diagonalized efficiently by using standard available libraries (like NAG) to 
generate accurate eigenvalues and eigenfunctions, for \emph{both} low as well as higher levels. 
Another notable feature is that this nonlinear optimal discretization maintains similar accuracy 
at \emph{both} smaller and larger distances with significantly smaller number of points (all 
calculations done with 250 radial points), than those needed in usual finite-difference or finite 
element methods but at the same time promises faster convergence. A convergence criterion of 
10$^{-5}$ and 10$^{-6}$ a.u. was imposed for the potential and energy to obtain all the results 
presented here. By performing a series of test calculations, a consistent set of GPS parameters 
were chosen ($\alpha$=25, N=250, $r_{max} \! = \! 200$ a.u.) which produced ``stable'' converged 
results. 

\begingroup
\squeezetable
\begin{table}
\caption {\label{tab:table1} Calculated total energies of ground and excited states of Li$^-$, 
Be$^-$ along with the literature data. Numbers in the parentheses denote absolute per cent 
deviations. See text for details.} 
\begin{ruledtabular}
\begin{tabular}{llllll}
Ion     &  State  & \multicolumn{4}{c}{$-$E(a.u.)}                  \\
\cline{3-6} 
        &         & \multicolumn{2}{c}{X-only}  &  \multicolumn{2}{c}{XC}       \\
\cline{3-4}  \cline{5-6}
        &                       &  This work &  Ref.     &   This work  &  Ref.  \\
\hline 
Li$^-$ & [He]2s$^2$ $^1$S$^e$   & 7.4278(0.005)   & 7.4282\footnotemark[1]     
                                & 7.4984(0.009)   & 7.4553\footnotemark[2],7.5008\footnotemark[1],
                                                    7.4991\footnotemark[3]        \\ 
       & 1s2s2p$^2$ $^5$P$^e$   & 5.3640(0.006)   & 5.3643\footnotemark[3] 
                                & 5.3925(0.171)   & 5.3866\footnotemark[4]$^,$\footnotemark[5],
                                                    5.3833\footnotemark[3],5.3865\footnotemark[6],
                                                    5.3863\footnotemark[11] \\
       & 1s2p$^3$ $^5$S$^o$     & 5.2223(0.004)   & 5.2225\footnotemark[3]
                                & 5.2608(0.137)   & 5.2561\footnotemark[4]$^,$\footnotemark[5],
                                                    5.2536\footnotemark[3],5.2560\footnotemark[6],
                                                    5.2558\footnotemark[11] \\
        & 1s2s2p3p $^5$P$^e$    & 5.3289          &   
                                & 5.3683(0.007)   & 5.3679\footnotemark[5]                      \\
Be$^-$ & [He]2s2p$^2$ $^4$P$^e$ & 14.5078(0.008)  & 14.5090\footnotemark[7] 
                                & 14.5806(0.062)  & 14.5779\footnotemark[8],14.5716\footnotemark[3],
                                                    14.5708\footnotemark[7],14.5769\footnotemark[10]\\
        & [He]2p$^3$ $^4$S$^o$  & 14.3272(0.002)  & 14.3275\footnotemark[7]
                                & 14.4081(0.049)  & 14.4063\footnotemark[8],14.4010\footnotemark[3],
                                                    14.4002\footnotemark[7]  \\
        & 1s2s2p$^3$ $^6$S$^o$  & 10.4279(0.009)  & 10.4288\footnotemark[7]
                                & 10.4758(0.092)  & 10.4662\footnotemark[3], 10.4615\footnotemark[7],
                                                    10.4711\footnotemark[9]  \\
\end{tabular}
\end{ruledtabular}
\begin{tabbing}
$^{\mathrm{a}}$Ref. \cite{fischer93}.   \= $^{\mathrm{b}}${Ref. \cite{weiss68}.} \=
$^{\mathrm{c}}${Ref. \cite{galvez06}.}  \= $^{\mathrm{d}}${Ref. \cite{yang95}.}  \=
$^{\mathrm{e}}${Ref. \cite{wang06}.}    \= $^{\mathrm{f}}${Ref. \cite{bunge80}.} \= 
$^{\mathrm{g}}${Ref. \cite{beck81}.}    \= $^{\mathrm{h}}${Ref. \cite{hsu95}.}   \= 
$^{\mathrm{i}}${Ref. \cite{hsu95a}.}    \= $^{\mathrm{j}}${Ref. \cite{bunge86}.} \= 
$^{\mathrm{k}}${Ref. \cite{fischer90}.} \= 
\end{tabbing}
\end{table}
\endgroup

\section{Results and Discussion}
Table I reports our calculated density functional term energies for ground and excited states of 
Li$^-$ and Be$^-$. We consider 4 states for the former, \emph{viz.,} [He]2s$^2$ $^1$S$^e$, 
1s2s2p$^2$ $^5$P$^e$, [He]2p$^3$ $^5$S$^o$, 1s2s2p3p $^5$P$^e$; and 3 for the latter,
[He]2s2p$^2$ $^4$P$^e$, [He]2p$^3$ $^4$S$^o$, 1s2s2p$^3$ $^6$S$^o$. For each of these states, two 
sets of energies are given; \emph{viz.,} X-only (noncorrelated) and XC (correlated). All of these 
states have been quite extensively studied, except the core-excited even-parity 1s2s2p3p $^5$P of 
Li$^-$, which has been reported only lately. The latter is also the only case for which HF result
is unavailable. Existing literature data are given for comparison. To put results in proper 
perspective, in parentheses the respective per cent deviations are given; for X-only case these are
relative to the lone literature results in column 4; for XC case, these are with respect to the 
recent variational Monte Carlo (VMC) \cite{galvez06} values except for the 4th states of Li$^-$, 
where such a result does not exist; and this is given in reference to the recent saddle-point 
calculation of \cite{wang06}. In VMC approach the authors employed explicitly correlated wave functions 
having a Jastraw factor and a multideterminant model wave function to account for the dynamical and 
nondynamical correlation effects respectively. Present X-only result of ground state of Li$^-$ is 
higher from the accurate HF calculation of \cite{fischer93} by only 0.0004 a.u. Our XC energy value 
is in fairly good agreement (slightly above) with the accurate correlated results reported through 
MCHF-$n$ expansion considering all expansions \cite{fischer93}, as well as the VMC method 
\cite{galvez06}. However the earlier result of \cite{weiss68} seems to be in considerable 
disagreement with these two and ours. The present X-only results for the next core-excited high-spin 
even-parity $^5$P$^e$ and odd-parity $^5$S$^o$ states of Li$^-$ again show excellent agreement with 
the recent HF estimates of \cite{galvez06}, while the XC energies match well with the VMC 
\cite{galvez06}, extensive CI \cite{bunge80}, variational multiconfiguration calculation 
\cite{yang95}, saddle-point \cite{wang06}, MCHF \cite{fischer90}, etc. Note that the present XC 
energies for these two states are lower than all of these reference results by 0.171 and 0.137\%, 
respectively and constitute the two instances giving maximum deviations in our calculation. Since 
our X-only results are virtually of HF quality, this overestimation is presumably caused by the 
correlation potential employed. It is worth noting here that even though a KS equation is solved
with the work-function exchange and LYP correlation potential, the procedure is \emph{not} subject 
to a variational bound \cite{roy97,roy98}. We note that some of these correlated calculations are 
highly elaborate and extensive; for example, \cite{yang95} used a 45 angular component 1004-term 
wave function, \cite{bunge80} used a 320-term CI, \cite{wang06} used 7-50 angular spin components 
and 541--1298 linear parameters for the former, and it is gratifying that current results exhibit a 
rather small and acceptable discrepancy. The next even-parity $^5$P$^e$ state has been investigated 
only recently \cite{wang06}, and our energy shows very good matching with theirs. They used a 
saddle-point restricted variational method with accurate multiconfigurational wave functions built 
from STO basis sets. As already mentioned, Be$^-$ has no [He]2s$^2$2p $^2$P ground state; and 
three metastable bound states found in the discrete spectrum are given in this table. [He]2s2p$^2$ 
$^4$P$^e$, [He]2p$^3$ $^4$S$^o$ and 1s2s2p$^3$ $^6$S$^o$ states lie below the Be 1s$^2$2s2p 
$^3$P, 1s$^2$2p$^2$ $^3$P and 1s2s2p$^2$ $^5$P excited states. The X-only energies again 
matching excellently with the HF results \cite{beck81}. A large number of accurate and sophisticated 
theoretical results exist for the correlated case. For example CI calculation of \cite{beck81} 
included single, double, triple and quadrupole subshell excitations. The first two states have also 
been studied through a method of full core plus correlation and restricted variation approach 
\cite{hsu95}. Our XC energies show better agreements with literature results in this case than those
for Li$^-$; however once again falling below the reference values for all the three states as in 
Li$^-$. Also a combined Rayleigh-Ritz and method of restricted variation result exists for the 
$^6$S$^o$ state \cite{hsu95a}.

Next, two calculated transition wavelengths (in nm) of Li$^-$ 1s2s2p$^2$ $^5$P$^e$ $\rightarrow$ 
1s2p$^3$ $^5$S$^o$ and Be$^-$ [He]2s2p$^2$ $^4$P$^e$ $\rightarrow$ [He]2p$^3$ $^4$S$^o$ are 
reported in Table II. A variety of theoretical and experimental results have been published, 
which are given for comparison. Note that the latter includes relativistic effects, whereas 
present calculation is nonrelativistic. Generally there seems to be good agreement for these quantities 
among the literature values. The present results are lower than the experimental values of
\cite{berry70} and \cite{gaardsted89} by 0.891\% and 0.438\% respectively for Li$^-$ and Be$^-$. 
In parentheses we give the corresponding wavelengths calculated from the X-only results. As 
expected they show considerable difference.  

\begingroup
\squeezetable
\begin{table}
\caption {\label{tab:table2}Comparison of transition wavelengths for Li$^-$, Be$^-$ with 
literature data. 1 a.u.=27.2113834 eV. $\hbar$c=197.3269602 eV nm. X-only values are enclosed in
parentheses.} 
\begin{ruledtabular}
\begin{tabular}{llll}
    &   \multicolumn{3}{c}{Wavelength (nm)} \\
\cline{2-4} 
Transition &  This work    &  Other theory &  Experiment \\  \hline
Li$^-$ 1s2s2p$^2$ $^5$P$^e$ $\rightarrow$ 1s2p$^3$ $^5$S$^o$        &   345.96
  & 346.06\footnotemark[1],349.12\footnotemark[2],349.0\footnotemark[3],348.98\footnotemark[4]
  & 349.07\footnotemark[5],349.0\footnotemark[6] \\ 
     &    (321.55)   &    &                      \\
Be$^-$ [He] 2s2p$^2$ $^4$P$^e$ $\rightarrow$ [He] 2p$^3$ $^4$S$^o$  &   264.14
  & 267.1\footnotemark[7],265.4\footnotemark[8],265.370\footnotemark[9],265.32\footnotemark[10],
    265.04\footnotemark[11]
  & 265.301\footnotemark[12],265.318\footnotemark[13],265.331\footnotemark[14]  \\
  &         (252.29)        &       &                     \\
\end{tabular}
\end{ruledtabular}
\begin{tabbing}
$^{\mathrm{a}}$Ref. \cite{wang06}.     \= $^{\mathrm{b}}${Ref. \cite{yang95}.} \=
$^{\mathrm{c}}${Ref. \cite{beck83}.}   \= $^{\mathrm{d}}${Ref. \cite{bunge80a}.}  \=
$^{\mathrm{e}}${Ref. \cite{berry70}.}  \= $^{\mathrm{f}}${Ref. \cite{mannervik80}.} \= 
$^{\mathrm{g}}${Ref. \cite{beck81}.}   \\ 
$^{\mathrm{h}}${Ref. \cite{beck84a}.}     \=
$^{\mathrm{i}}${Ref. \cite{hsu95}.}    \= $^{\mathrm{j}}${Ref. \cite{fischer90}.}   \= 
$^{\mathrm{k}}${Ref. \cite{olsen94}.}  \= $^{\mathrm{l}}${Ref. \cite{gaardsted89}.} \= 
$^{\mathrm{m}}${Ref. \cite{kristensen95}.} \= $^{\mathrm{n}}${Ref. \cite{andersen96}.} 
\end{tabbing}
\end{table}
\endgroup

\begin{figure}
\begin{minipage}[c]{0.40\textwidth}
\centering
\includegraphics[scale=0.38]{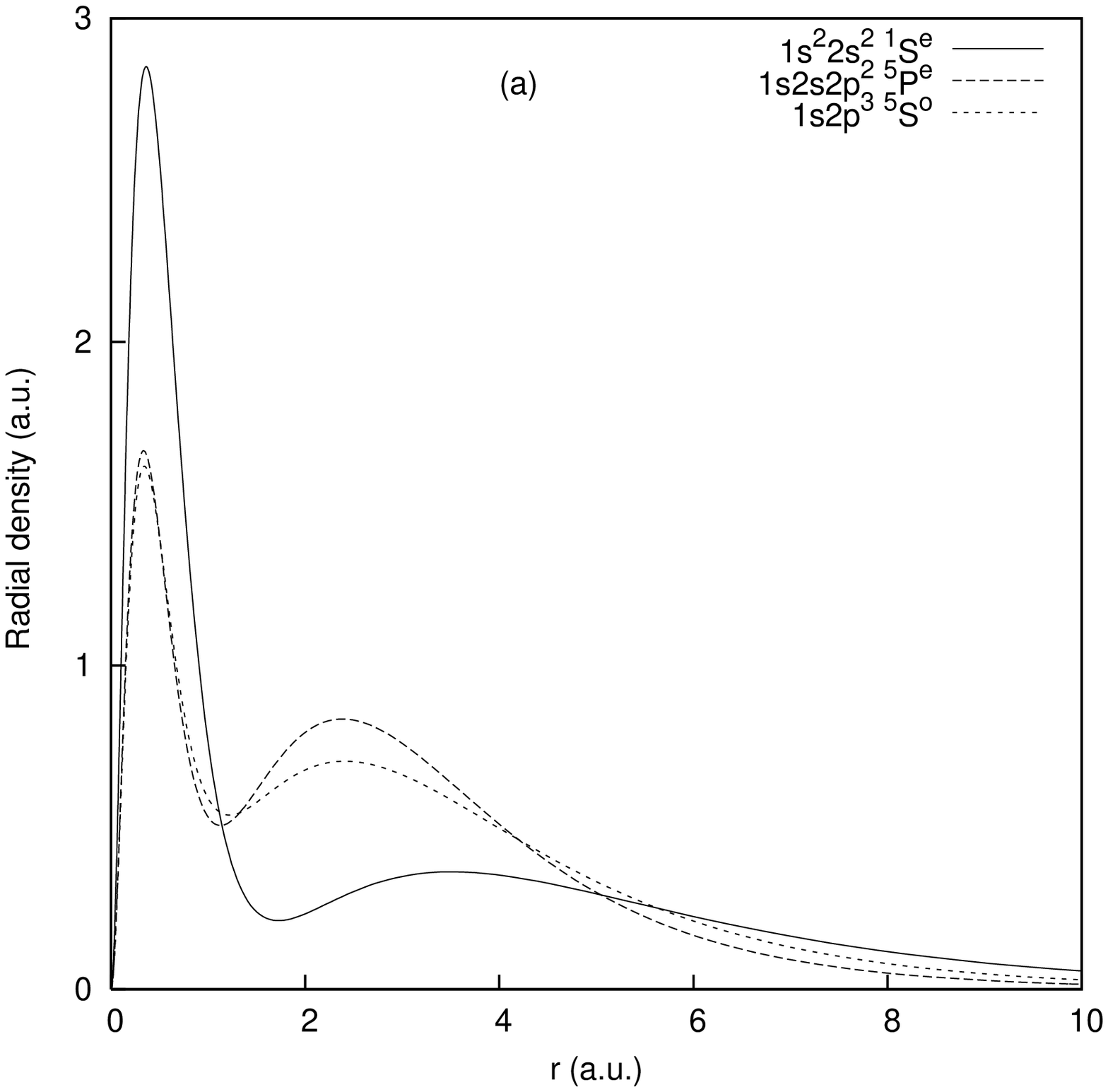}
\end{minipage}%
\hspace{0.3in}
\begin{minipage}[c]{0.40\textwidth}
\centering
\includegraphics[scale=0.38]{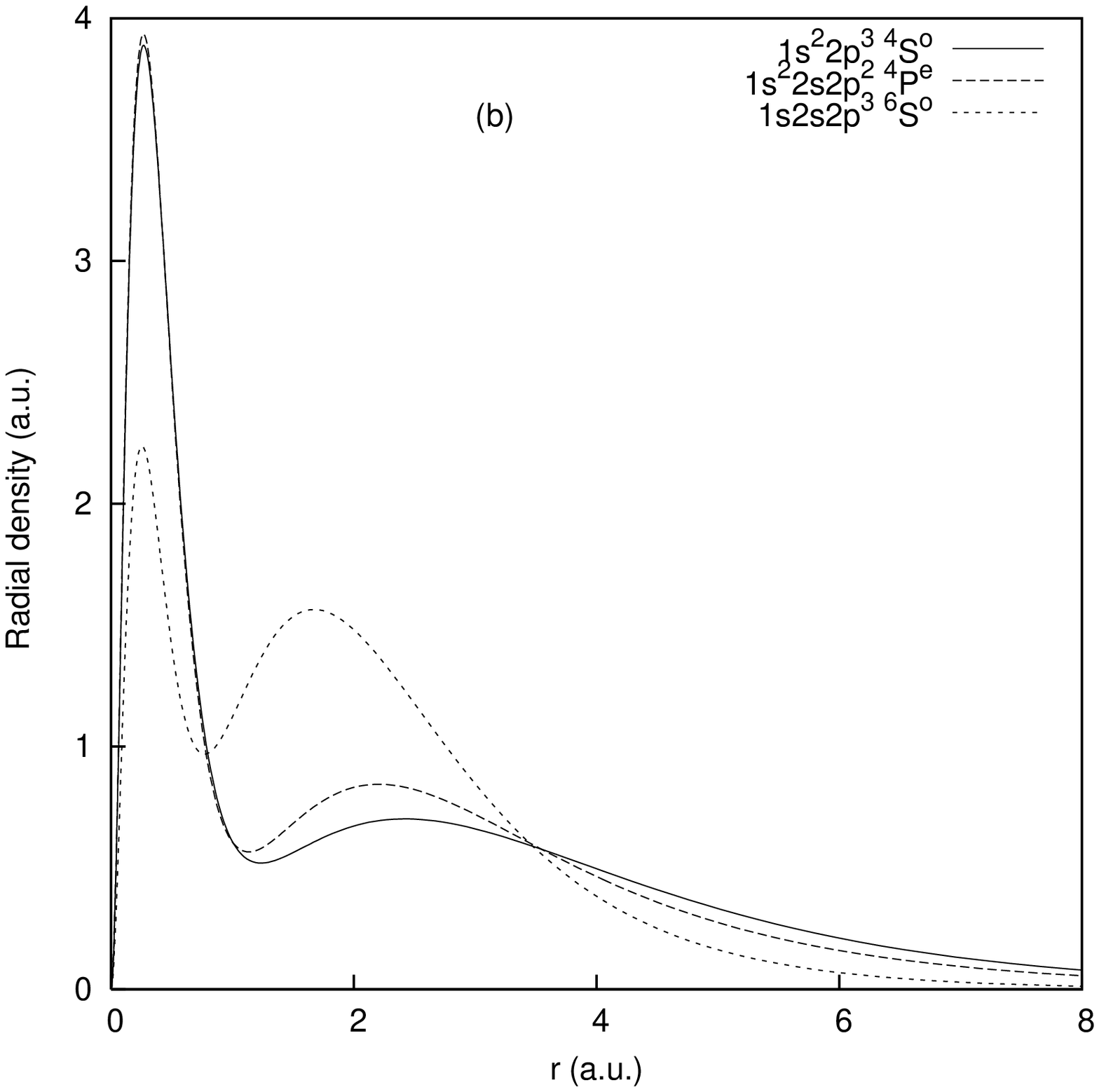}
\end{minipage}%
\caption{Radial density plots for several states of (a) Li$^-$ and (b) Be$^-$.}
\end{figure}
Now Fig. 1(a) depicts the single-particle radial density plots for Li$^-$ ground state [He]2s$^2$ 
$^1$S$^e$, and two excited states 1s2s2p$^2$ $^5$P$^e$, 1s2p$^3$ $^5$S$^o$. Close to the nucleus, 
the latter two states making a family of similar radial densities, possess considerably smaller 
charge density than the ground state, mainly due to the presence of two core 1s electrons in the 
ground state and one 1s electron in the latter two. However this scenario reverses in the region
enclosed between the first minimum and up to $r\!=\!4$ a.u., where the $^5$P$^e$ and $^5$S$^o$ 
states show larger values than the $^1$S$^e$ state. The two excited states also appear to separate 
out in this intermediate region with $^5$P$^e$ showing larger peak value than the $^5$S$^o$. After 
that, at large distances, all these three states show similar behavior with the ground state having 
greater charge density and decaying rather slowly than the two excited states; $^5$P$^e$ dying out 
fastest. Similar radial density plots for all three excited states of Be$^-$ are given in Fig. 1(b). 
In the vicinity of nucleus, the $^4$P$^e$, $^4$S$^o$ states show similar behavior and have much 
larger charge densities compared to the $^6$S$^o$ state, again presumably because the former two 
have two core 1s electrons while the latter has only one. At the intermediate distance after the 
first minimum and up to $r\!=\!3.5$ a.u., this situation changes with $^6$S$^o$ having the largest 
charge density. Also the $^4$S$^o$, $^4$P$^e$ states branch out in this region with the latter 
showing slightly larger peak value than the former. After that at larger distances, all the three 
states decay in a similar pattern with $^4$S$^o$ having higher values and oozing out slowly than the 
other two and $^6$S$^o$ decaying out first. These behaviors in electron density are also reflected 
in the selective radial expectation moments $\langle r^{-2} \rangle$, $\langle r^{-1} \rangle$, 
$\langle r^{1} \rangle$ and $\langle r^{2} \rangle$ values of these 6 states of Li$^-$, Be$^-$ as
compiled in Table III. Some of these quantities have been recently reported in the VMC 
calculation of \cite{galvez06}, quoted in parentheses for comparison. Generally there is good 
agreement between our results with them, especially for the negative moments and discrepancy tends 
to increase for the positive moments. However the general trend is similar as theirs. The negative 
expectation values of Li$^-$ are reduced for both the excited states ($^5$P$^e$, $^5$S$^e$) compared 
to the ground state, and having closer values to each other, due to the close resemblance of their 
corresponding radial densities close to the nucleus. $^5$P$^e$ state with lowest energy showing 
larger $\langle r^{-2} \rangle $ and $\langle r^{-1} \rangle$ than the $^5$S$^o$. The positive 
moments however correspond more to the charge densities at intermediate-to-large distances, and show
complicated behavior. For Be$^-$, the negative moments are smaller for $^6$S$^o$ term compared to 
the quartet S and P terms, with the latter two having similar values, as they have the same [He] 
core. As in Li$^-$, the P state (with lower energy) gives higher values of the negative moments 
compared to the S state.

\begingroup
\squeezetable
\begin{table}
\caption {\label{tab:table3}Calculated expectation values in a.u. Numbers in the parentheses are
taken from \cite{galvez06}.}
\begin{ruledtabular}
\begin{tabular}{llllll}
 Ion  &  State & $\langle r^{-2}\rangle$ & $\langle r^{-1}\rangle$ 
               & $\langle r \rangle$ & $\langle r^2 \rangle$  \\    \hline
Li$^-$  & [He]2s$^2$ $^1$S$^e$    & 30.458(30.19)    &   5.920(5.8977)   
                                  & 11.846(11.718)   &  73.713(71.30)      \\
        & 1s2s2p$^2$ $^5$P$^e$    & 19.022(19.04)    &   4.173(4.1548) 
                                  & 11.933(11.7410)  &  53.197(57.726)     \\  
        & 1s2p$^3$ $^5$S$^o$      & 18.325(18.23)    &  4.094(4.0475)   
                                  & 12.148(12.9592)  &  70.686(72.73)      \\
Be$^-$  & [He]2s2p$^2$ $^4$P$^e$  &  56.927(56.7)    &  8.622(8.5525)     
                                  & 11.131(12.478)   &  62.927(66.86)      \\ 
        & [He]2p$^3$ $^4$S$^o$    &  56.083(55.7)    &  8.485(8.4140)     
                                  &  12.173(14.135)  &  82.928(89.97)      \\
        & 1s2s2p$^3$ $^6$S$^o$    &  34.468(34.6)    &  6.276(6.2137)  
                                  & 10.038(10.5831)  &  30.512(34.785)      \\ 
\end{tabular} 
\end{ruledtabular}
\end{table}
\endgroup

\section{Conclusion}
A density-based formalism was employed for the \emph{first} time to calculate the state energies, radial 
density as well as selected density moments of three excited states of Li$^-$ and Be$^-$ in addition 
to the ground state of the former. A work-function-based exchange potential plus the nonlinear LYP
correlation potential was used, while the resulting KS equation was numerically solved by an 
optimal, nonuniform, spatial GPS discretization scheme accurately. Additionally two transition
wavelengths of Li$^-$, Be$^-$ were computed. Comparison with the existing theoretical as well as 
experimental data were made. For all practical purposes our X-only results are of HF quality and 
with correlation included, they produce excellent agreement with the literature data. The observed 
absolute per cent deviation in the term energies between our XC calculations and the correlated 
results in the literature remains within 0.007-0.171\% only, and is mostly attributable to our 
current inadequate knowledge of the delicate correlation effects in these strongly-correlated 
systems, besides the fact that the presently employed LYP potential was designed primarily for the 
ground states. Construction of the \emph{exact} many-body correlation functional has been one of 
the most forefront areas of DFT ever since its rejuvenation in the works of Kohn-Sham in 1964 and 
unfortunately still remains one of the most important unresolved issues in DFT. However, even though 
its \emph{exact} form remains elusive, many accurate 
forms have been proposed over the decades suitable for different processes in atoms, molecules, 
solids. Note that this is a single-determinantal approach and is far more simpler to implement 
compared to the benchmark sophisticated and elaborate CI or multiconfiguration type formalisms
within the traditional wave function based methodologies; yet as our results indicate, this can 
lead to quite good-quality results, which of course is the appeal and usefulness of DFT-based 
methods. At this stage, we note that, success of such a relatively simpler method on the present and 
previous atomic 
excited states done before, e.g., in \cite{roy97, roy98, roy02, roy04}, may encourage one to explore 
its validity and applicability for other physical and chemical systems such as molecules. While this
could lead to an attractive and direct route towards molecular excited states, as discussed earlier,
(see for example, \cite{roy02} and references therein), a straightforward extension is difficult; as it 
relies on the tacit assumption of spherical symmetry in the density, which is not the case for 
molecules. Finally, while the LYP correlation potential employed has provided rather excellent results 
for the present inherently correlated systems and in past studies, nevertheless it is unclear whether
the non-dynamical correlation contribution is described by LYP properly or for these states such 
contributions are small. No definitive answer is possible and more detailed and sophisticated studies 
would be required for this. In conclusion, we have 
proposed an accurate and reliable method for the excited states of negative atoms within the 
framework of DFT.

\begin{acknowledgments}
AKR thanks Professors D.~Neuhauser, S.~I.~Chu and B.~M.~Deb for encouragement, support and useful 
discussions. It is a pleasure to thank Dr.~E.~Proynov for his comments. He acknowledges the 
warm hospitality provided by the Univ. of California, Los angeles, CA, USA. It is a pleasure to thank
the anonymous referee for comments.
\end{acknowledgments}

\end{document}